\documentclass[aps,prb,preprint,twocolumn,showpacs,floatfix,10pt]{revtex4-1}
\usepackage{amsfonts}
\usepackage{amsmath}
\usepackage{amssymb}
\usepackage{graphicx}
\usepackage{epstopdf}
\usepackage{float}%
\begin{document}
\title{Electron transport through a diatomic molecule}
\author{$^{^{\dag}}$M. Imran}
\affiliation{Department of Physics, Quaid-i-Azam University, 49000 Islamabad, Pakistan.}


\begin{abstract}
Electron transport through a diatomic molecular tunnel junction shows wave like interference phenomenon. By using Keldysh non-equilibrium Green's function (NEGF) theory, we have explicitly presented current and differential conductance calculation for a diatomic molecular and two isolated atoms (two atoms having zero hybridization between their energy orbital) tunnel junctions. In case of a diatomic molecular tunnel junction, Green's function propagators entering into current and differential conductance formula interfere constructively for a molecular anti-bonding state and destructively for bonding state. Consequently, conductance through a molecular bonding state is suppressed, and to conserve current, conductance through anti-bonding state is enhanced. Therefore, current steps and differential conductance peaks amplitude show asymmetric correspondence between molecular bonding and anti-bonding states.  Interestingly, for a diatomic molecule, comprising of two atoms of same energy level, these propagators interfere completely destructively for molecular bonding state and constructively for molecular anti-bonding state. Hence under such condition, a single step or a single peak is shown up in current versus voltage or differential conductance versus voltage studies.

\end{abstract}

\startpage{1}
\endpage{2}
\maketitle
\section{Introduction}
Aptitude of semiconductor technology towards miniaturisation of electronic
devices has lead to emergence of nano and molecular electronics. Now the work
of nano scientists is to control electron transport at such minute scale. As
the size of the nano devices is comparable to the wavelength of the current
carriers so quantum features are playing dominant role in these devices. A
theoretical model of electron transport through a single molecule was first presented
by A.Aviram and Mark.A.Ratner \cite{1,2}. Since then there has been great
interest both from experimental and theoretical groups for the progress of
molecular electronics \cite{3,4,5}. A single molecule could be used as
electronic mixers \cite{6}, switches \cite{7,8}, and rectifiers \cite{9}. With an
eye on its applications, it is expected that the understanding of quantum electron
transport at the molecular scale is a key step towards practical usage of molecular
devices \cite{10}. Experiments on conduction through molecular tunnel junctions
are becoming more common, for discussion see Ref \cite{11,12} and references therein. On the
experimental front, the most common methods of contacting individual molecule
are scanning tunnelling microscope tip and mechanically controlled
break junctions \cite{13,14}. Early experiments in the field of molecular electronics have focused on the absolute
conductance and its dependence on wire length, molecular
structure, and temperature \cite{29}. From the theoretical point of view, investigating
electron transport in an electrically contacted molecule is a challenging
problem. Most of the formal theoretical work on transport through molecular
tunnel junction have relied on the Generalized master equation's theory \cite{15,16} and the non equilibrium Green's function (NEGF)
theory \cite{17,18}. By using NEGF theory a plethora of physical phenomena like
coulomb blockade \cite{19}, Kondo effect \cite{20} and vibrations
effect \cite{21} are already reported.

 Electron transport through a molecular tunnel junction has very non-intuitive
characteristics. Many experimentalists and theoreticians of molecular science
refer to hydrogen molecular transport to be explained by transport through a
single channel \cite{23,24,25,26}, instead of the presence of hydrogen molecular bonding and anti-bonding states, as conductance channels.
On the basis of the first-principle calculations, it has been already reported that the transmission probability through hydrogen molecular tunnel junction becomes exactly one for molecular anti-bonding state \cite{27}. Apart from the hydrogen molecule, even more complex molecules
show asymmetry in conductance through highest occupied molecular orbital
(HOMO) and lowest unoccupied molecular orbital (LUMO) \cite{28}.

In this study we discuss a diatomic molecular tunnel junction and contrast it with two isolated atoms tunnel junction. The results obtained by the present theoretical formulation is valid even for artificial diatomic molecule (two coupled quantum dots \cite{30}). By using NEGF theory, we have explicitly shown the current and differential conductance of the under discussion system. In the presence of the hybridization between energy orbital of the two atoms, a diatomic molecular bonding and anti-bonding states are formed. The Green's function propagators entering into current and differential conductance interfere constructively for two isolated atoms tunnel junction, where as, for a diatomic molecular tunnel junction, they interfere constructively for an anti-bonding state and destructively for a bonding state. Therefore, the normalized transmission probability function for a diatomic molecular tunnel junction shows asymmetric correspondence between molecular bonding and anti-bonding states. For a diatomic molecule comprising of two atoms of the same energy level the normalized transmission probability becomes one for molecular anti-bonding state and zero for molecular bonding state. This clarifies the asymmetric correspondence in current steps and differential conductance peaks amplitude for a diatomic molecular tunnel junction. 

This paper is organized as follows. In section II a diatomic molecular tunnel junction model Hamiltonian, current and differential conductance are presented. In section III numerical results are discussed. Conclusions are made in section IV.
\section{Theoretical Modelling}
In modelling Hamiltonian of a diatomic molecule employed between electrodes, we have
considered electrodes as charge carriers bath and charge carriers wave function
hybridize between electrodes and i-th atom of a diatomic molecule. The Hamiltonian of the under discussion
mesoscopic system is,%
\begin{equation}
\begin{split}
H= \sum_{i=1}^{2} \epsilon_{i}d_{i}^{\dag}d_{i}+\sum_{i,j=1,i\neq j}^{2}\tau
d_{i}^{\dag}d_{j}+\\
\sum_{k\nu i}(T_{\nu ki}d_{i}^{\dag}c_{\nu k}+T_{\nu k
i}^{\dag}c_{\nu k}^{\dag}d_{i})+\sum_{k}\epsilon_{\nu k}c_{\nu k}^{\dag}c_{\nu k}
\label{1}%
\end{split}
\end{equation}
The  Hamiltonian is written under the approximation of linear combination
of atomic orbitals and in second quantized form \cite{18}. Here $d_{i}^{\dag
}(d_{i})$ represents creation
(annihilation) operator of electron on i-th atom, where as
second term refers to inter-atomic hybridization energy $\tau$. The $\epsilon_{i}$,$\epsilon_{\nu k}$, and $T_{\nu k}$ are
representing i-th atom, electrodes, and i-th atom and electrodes hybridization
energy, respectively. The operator $c_{\nu k }^{\dag}(c_{\nu k })$  is  electron creation (annihilation)
operator on the $\nu$ electrode. We have assumed for simplicity $\hbar=e=1$.

The current across a diatomic molecular tunnel junction is calculated by the time evolution of the occupation
number operator $n^{\nu}\left(  t\right)  =\sum_{k}{\epsilon}%
_{k}\left(  t\right) {c}_{\nu k}^{\dag}{c}_{\nu k}$ of the
$\nu$ electrode.%
\begin{equation}%
\begin{split}
I_{\nu}\left(  t\right)   &  =-\langle\frac{\partial}{\partial t}n^{\nu
}\left(  t\right)  \rangle=-i\langle\left[  H,n^{\nu}\left(  t\right)
\right]  \rangle\\
&  =2\operatorname{Re}[\sum_{ki}{T}_{\nu ki}Gc^{<}(t,t)]
\end{split}
\end{equation}

Here, $Gc^{<}(t,t)=i\left\langle
{c}_{\nu k}^{\dag}d_{i}\right\rangle $ represents electrode and a diatomic molecular coupled lesser Green's function. 
 A diatomic molecular and $\nu$ electrode coupled lesser Green's function can be de-coupled into a diatomic
molecular ($G_{ij}^{R,<}$) and $\nu$ electrode ($g_{\nu k}^{R,<}$) Green's function by exploiting analytic
continuation rule \cite{18}%
\begin{equation}%
\begin{split}
Gc^{<}(t,t^{\prime})  &  =\sum_{j\nu k}{T}_{\nu kj}^{\dag}\int
dt_{1}\left[  G_{ij}^{R}(t,t_{1})g_{\nu k}^{<}(t_{1},t^{\prime})\right. \\
&  \left.  +G_{ij}^{<}(t,t_{1})g_{\nu k}^{A}(t_{1},t^{\prime})\right]
\end{split}
\end{equation}

The $\nu$ electrode lesser and advanced Green's function are $g_{\nu k}^{A}(t,t^{\prime})=i\theta\left(t^{\prime}-t\right)Exp\left[
i\epsilon_{k}\left(  t-t^{\prime}\right)  \right]$ and $g_{\nu k}^{<}(t,t^{\prime})=if_{\nu}\left(  \epsilon_{k}\right)  Exp\left[
-i\epsilon_{k}\left(  t-t^{\prime}\right)  \right]$.\\
Here $f_{\nu
}\left(  \epsilon_{k}\right)  =%
\genfrac{}{}{0.5 pt}{0}{1}{Exp\left[  \beta\left(  \epsilon_{k}-\mu_{\nu}\right)
\right]  +1}%
$ is the Fermi-Dirac distribution function. Where as, $\beta=%
\genfrac{}{}{0.5 pt}{0}{1}{k_{B}T}%
$ gives inverse of the thermal energy, with Boltzmann's constant $k_{B}$
and temperature $T$. By employing wide band approximation
($\sum_{k}\longrightarrow D\int_{-\infty}^{\infty}d\epsilon_{k}$) and using electrode Green's function in Eq. (3),  the current from a diatomic molecular tunnel junction is given by the following relation.

\begin{equation}
{\scriptsize
\begin{split}
I_{\nu}\left(  t\right)   &  =-2\operatorname{Im}\left[  \sum_{ij}%
{\Gamma}_{\nu ij}\int dt_{1}\int\frac{d\epsilon_{k}}{2\pi}\right. \\
&  \left.  \left(  Exp\left[  i\epsilon_{k}\left(  t-t_{1}\right)  \right]
\right)  \left[  f_{\nu}\left(  \epsilon_{k}\right)  G_{ij}^{R}(t,t_{1}%
)+G_{ij}^{<}(t,t_{1})\right]  \right]  \text{.}%
\end{split}
}%
\end{equation}
Here ${\Gamma}_{\nu ij}=2\pi D{T}_{\nu ki}{T}_{\nu
kj}^{\dag}$.  By utilizing the conditions of current continuity and line width proportionality \cite{1} ${\Gamma}_{Lij}=\eta{\Gamma
}_{Rij}$, the current through a diatomic molecular tunnel junction is given as follows.%
\begin{equation} %
{\scriptsize
\begin{split}
I_{L}\left(  t\right)   &  =-\Gamma\sum_{ij}
\operatorname{Im}\left[
\int\frac{d\epsilon_{k}}{2\pi}\int dt_{1}\right. \\
&  \left.  \left[  f_{L}\left(  \epsilon_{k}\right)  -f_{R}\left(
\epsilon_{k}\right)  \right]  Exp\left[  i\epsilon_{k}\left(  t-t_{1}\right)
\right]  G_{ij}^{R}(t,t_{1})\right]  %
\end{split}
} %
\end{equation}
Here, $\Gamma=\Gamma_{Rij}=\Gamma_{Lij}$.
Now we use Dyson's equation to find a diatomic molecular-system total retarded
Green's function $G_{ij}^{R}$.%
\begin{equation}%
{\scriptsize
\begin{split}
G_{ij}^{R}(t,t^{\prime})  &  =g_{ij}^{R}(t,t^{\prime})+\sum_{\nu k}%
{T}_{\nu ki}{T}_{\nu kj}^{\dag}\\
&  \int dt_{1}\int dt_{2}g_{ij}^{R}(t,t_{1})g_{\nu k}^{R}(t_{1},t_{2})G_{ij}%
^{R}(t_{2},t^{\prime})
\end{split}
} %
\end{equation}
Here, $g_{ij}^{R}$ and $g_{\nu k} ^{R}$ represent a diatomic molecular and $\nu$ electrode retarded Green's functions, respectively. By
using electrode Green's function into Eq.(6), diatomic
molecular-system total retarded Green's function is simplified.%
\begin{equation}%
\begin{split}
G_{ij}^{R}(t,t^{\prime})  &  =g_{ij}^{R}(t,t^{\prime})  Exp\left[  -\Gamma  \left(  t-t^{\prime}\right)  \right]  %
\end{split}
\end{equation}
Now a diatomic molecular retarded Green's $g_{ij}^{R}$ function is found by
the equation of motion technique.
\begin{equation}
i\dfrac{\partial}{\partial t}[Exp(i\epsilon_{_{1}}t)d_{1}(t)]=\tau
Exp(i\epsilon_{_{1}}t)d_{2}(t) \label{2}%
\end{equation}
\begin{equation}
i\dfrac{\partial}{\partial t}[Exp(i\epsilon_{2}t)d_{2}(t)]=\tau Exp(i\epsilon
_{2}t)d_{1}(t) \label{3}%
\end{equation}
By defining $F_{1}(t)\equiv Exp(i\epsilon_{_{1}}t)d_{1}(t)$ \\and $F_{2}%
(t)\equiv Exp(i\epsilon_{2}t)d_{2}(t)$
\begin{equation}
\frac{i}{\tau}Exp(i\Delta t)\dfrac{\partial}{\partial t}F_{1}(t)=F_{2}(t)
\label{6}%
\end{equation}
\begin{equation}
\frac{i}{\tau}Exp(-i\Delta t)\dfrac{\partial}{\partial t}F_{2}(t)=F_{1}(t)
\label{7}%
\end{equation}
Here $\Delta=(\epsilon_{_{2}}-\epsilon_{_{1}})$. For de-coupling Eqs. 10 and 11, we operate
$i\dfrac{\partial}{\partial t}$ on above two equations.
\begin{equation}
\dfrac{\partial^{2}F_{1}
}{\partial t^{2}}+i\Delta\dfrac{\partial F_{1}}{\partial t}+\tau^{2}F_{1}=0 \label{8}%
\end{equation}
\begin{equation}
\dfrac{\partial^{2}F_{2}
}{\partial t^{2}}-i\Delta\dfrac{\partial F_{2}}{\partial t}+\tau^{2}F_{2}=0 \label{8}%
\end{equation}
By solving above two differential equations, we find time evolution of the annihilation operators of the diatomic molecule.
{\footnotesize
\begin{align}
d_{1}(t)  &  =\frac{1}{2\gamma}[\{(\gamma+\Delta)d_{1}(0)-2\tau d_{2}%
(0)\}Exp(-i\epsilon_{b}t)+\}\nonumber\\
&  \{(\gamma-\Delta)d_{1}(0)+2\tau d_{2}(0)\}Exp(-i\epsilon_{a}t)\}]\label{10}
\end{align}
}
{\footnotesize
\begin{align}
d_{2}(t)  &  =\frac{1}{2\gamma}[\{(\gamma-\Delta)d_{2}(0)-2\tau d_{1}%
(0)\}Exp(-i\epsilon_{b}t)+\nonumber\\
&  \{(\gamma+\Delta)d_{2}(0)+2\tau d_{1}(0)\}Exp(-i\epsilon_{a}t)\}]\label{11}
\end{align}
}
Here $\gamma=\sqrt{\Delta^{2}+4\tau^{2}}$ , $\epsilon
_{b}=\dfrac{(\epsilon_{1}+\epsilon_{2}-\gamma)}{2}$ \ and $\epsilon_{a}%
=\dfrac{(\epsilon_{1}+\epsilon_{2}+\gamma)}{2}$. Where as
$\epsilon_{b}(\epsilon_{a})$ represents molecular bonding (molecular
anti-bonding) energy.
Similarly,
{\footnotesize
\begin{align}
d_{1}^{\dag}(t)  &  =\frac{1}{2\gamma}[\{(\gamma+\Delta)d_{1}^{\dag}(0)-2\tau
d_{2}^{\dag}(0)\}Exp(i\epsilon_{b}t)+\nonumber\\
&  \{(\gamma-\Delta)d_{1}^{\dag}(0)+2\tau d_{2}^{\dag}(0)\}Exp(i\epsilon
_{a}t)\}]\label{12}
\end{align}}
{\footnotesize
\begin{align}
d_{2}^{\dag}(t)  &  =\frac{1}{2\gamma}[\{(\gamma-\Delta)d_{2}^{\dag}(0)-2\tau
d_{1}^{\dag}(0)\}Exp(i\epsilon_{b}t)+\nonumber\\
&  \{(\gamma+\Delta)d_{2}^{\dag}(0)+2\tau d_{1}^{\dag}(0)\}Exp(i\epsilon
_{a}t)\}]\label{13}
\end{align}} %
Now a diatomic molecular retarded Green's function $g_{ij}^{R}$ is found by the following relation.
\begin{equation}
g_{ij}^{R}\left(  t,t^{\prime}\right)  \equiv-i\theta(t-t^{\prime
})\left\langle [d_{i}(t)\text{ \ }d_{j}^{\dagger}\left(  t^{\prime}\right)
]\right\rangle \label{15}%
\end{equation}

By utilizing Eqs. 14-18 and using anti-commutation algebra $\left[  d_{i}(0)\text{ }d_{j}^{\dag}(0)\right]
=\delta_{ij}$, a diatomic molecular-system total retarded Green's function is found.
{\small
\begin{equation}
G_{11}^{R}\left(  \epsilon\right)  =\frac{1}{2\gamma}[\dfrac{(\gamma+\Delta
)}{(\epsilon-\epsilon_{b}+i\Gamma)}+\dfrac{(\gamma-\Delta)}{(\epsilon
-\epsilon_{a}+i\Gamma)}] \nonumber%
\end{equation}}%
{\small
\begin{equation}
G_{12}^{R}\left(  \epsilon\right)  =\dfrac{\tau}{\gamma}[\dfrac{1}%
{(\epsilon-\epsilon_{a}+i\Gamma)}-\dfrac{1}{(\epsilon-\epsilon_{b}+i\Gamma)}]\nonumber
\end{equation}}%
{\small
\begin{equation}
G_{21}^{R}\left(  \epsilon\right)  =\dfrac{\tau}{\gamma}[\dfrac{1}%
{(\epsilon-\epsilon_{a}+i\Gamma)}-\dfrac{1}{(\epsilon-\epsilon_{b}+i\Gamma)}]\nonumber
\end{equation}}%
{\small
\begin{equation}
G_{22}^{R}\left(  \epsilon\right)  =\frac{1}{2\gamma}[\dfrac{(\gamma-\Delta
)}{(\epsilon-\epsilon_{b}+i\Gamma)}+\dfrac{(\gamma+\Delta)}{(\epsilon
-\epsilon_{a}+i\Gamma)}] \nonumber%
\end{equation}} %
Finally, current and differential conductance for a diatomic molecular tunnel junction are given by the following relations.
\begin{equation}
I=\int\frac{d\epsilon_{k}}{2\pi}T\left(
\epsilon_{k}\right)  \left[  f_{L}\left(  \epsilon_{k}\right)  -f_{R}\left(
\epsilon_{k}\right)  \right] %
\end{equation}%
and
\begin{equation}
\begin{aligned}
{\footnotesize
\begin{split}
\frac{dI}{dV}=\int\frac{d\epsilon_{k}}{2\pi
}T\left(  \epsilon_{k}\right)  \left[
\genfrac{}{}{0.5pt}{0}{Exp\left[  \beta\left(  \epsilon_{k}-\mu_{L}\right)
\right]  }{\left(  Exp\left[  \beta\left(  \epsilon_{k}-\mu_{L}\right)
\right]  +1\right)  ^{2}}%
+\right.\\\left.
\genfrac{}{}{0.5pt}{0}{Exp\left[  \beta\left(  \epsilon_{k}-\mu_{R}\right)
\right]  }{\left(  Exp\left[  \beta\left(  \epsilon_{k}-\mu_{R}\right)
\right]  +1\right)  ^{2}}%
\right]
\end{split}
}\nonumber
\end{aligned}
\end{equation}

with transmission function $  T\left(  \epsilon_{k}\right)   $

$T\left(  \epsilon_{k}\right)  =-\Gamma \sum_{ij}
\operatorname{Im}\left[  G_{ij}^{R}\left(  \epsilon_{k}\right)\right]$
\section{Numerical Results}
The comparative analysis of the electron transport between two isolated atoms tunnel junction and a diatomic molecular tunnel junction is discussed in this study. The quantum interference effects are usually destroyed in macroscopic devices. But in nano devices, where the wave length of the current carriers become comparable with the size of the device, the wave features are preserved. For two isolated atoms tunnel junction, current starts flowing as the Fermi energy level of the electrode equates with the energy level of the either atom. However, when the energy orbital of these atoms hybridises then a diatomic molecular bonding and anti-bonding states are formed.  Now current starts flowing as the Fermi energy level of the electrode equates with a diatomic molecular bonding or anti-bonding states.
\begin{figure}
 [b!]
 \begin{center}
 \includegraphics[
 height=4.5 cm,
 width=7 cm
 ]%
 {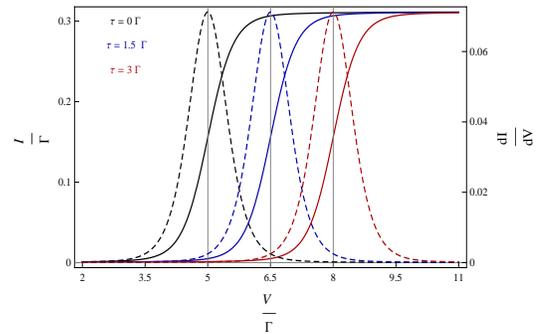}%
 \caption{(Color online) Current (solid lines) differential conductance (dashed lines). $\epsilon_{1}=\epsilon_{2}=5\Gamma$ and $\beta\Gamma=0.35$.}%
 \end{center}
 \end{figure}
In this study we have explicitly discussed that the electron transport through a diatomic molecular tunnel junction shows asymmetric correspondence in current steps and differential conductance peaks amplitude for molecular bonding and anti-bonding states.  More interestingly, for a diatomic molecular tunnel junction, where constituents of a diatomic molecular energy states are having same energy level then only a single step in current and a single peak in differential conductance is shown up \cite{27}. This could be seen in Fig. 1, where current and differential conductance functional relation with applied voltage is plotted.
 When two isolated atoms having the same energy level are placed between electrodes then only a single step in current and a single peak in differential conductance is shown up. While in the presence of finite hybridization between energy orbital of these atoms, a single step in current and a single peak in differential conductance is shown up at the anti-bonding energy level of the molecule. 
 \begin{figure}
 [pt]
 \begin{center}
 \includegraphics[
  height=4.5 cm,
  width=7 cm
     ]%
    {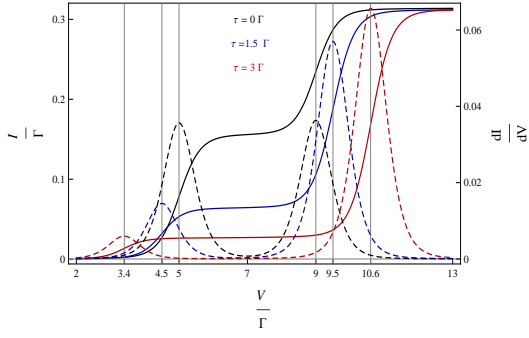}%
     \caption{(Color online) Current (solid lines) differential conductance (dashed lines). $\epsilon_{1}=5\Gamma$, $\epsilon_{2}=9\Gamma$, and $\beta\Gamma=0.35$.}%
    \end{center}
    \end{figure}
 \begin{figure}
 [pb]
 \begin{center}
 \includegraphics[
  height=4.5 cm,
  width=6 cm
 ]%
 {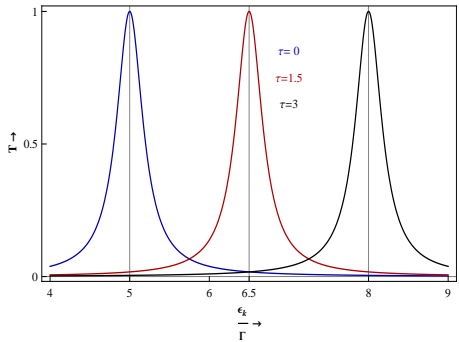}%
 \caption{(Color online) The normalized transmission probability. $\epsilon_{1}=\epsilon_{2}=5\Gamma$.}%
 \end{center}
 \end{figure}
Where as, no current step and differential conductance peak is shown up at the bonding energy level of the molecule.  The increase in hybridization between energy orbital of these atoms, induces shift between molecular bonding and anti-bonding energy levels. Consequently, current step and differential conductance peak too shift to the new molecular anti-bonding energy level.
  \begin{figure}
   [t!]
   \begin{center}
   \includegraphics[
 height=4.5 cm,
 width=6 cm
   ]%
   {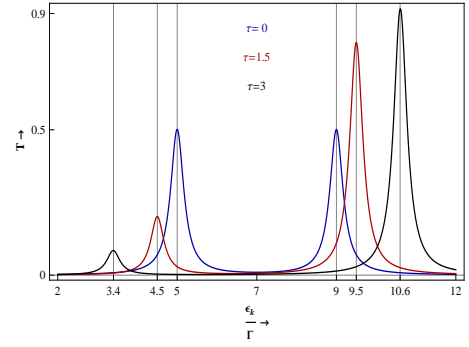}%
   \caption{(Color online) The normalized transmission probability. $\epsilon_{1}=5\Gamma$, and $\epsilon_{2}=9\Gamma$.}%
   \end{center}
   \end{figure}
   
The asymmetry in conductance through a diatomic molecular energy states could be more clearly visualised by studying electron transport through a diatomic molecule comprising of two atoms of different energy level. See Fig. 2.
In the absence of hybridization between energy orbital of the atoms, two symmetric steps and peaks are shown up in current and differential conductance, respectively. The presence of hybridization between energy orbital of these atoms results in formation of molecular bonding and anti-bonding states. The current step and differential conductance peak amplitude of molecular bonding state are suppressed and consequently these are enhanced for molecular anti-bonding state.

To investigate the asymmetry in conductance through molecular bonding and anti-bonding states, we discuss the normalized transmission probability. The normalized transmission probability $\dfrac{\Gamma^{2}}{(\epsilon_{k}-\epsilon_{0})^{2}+\Gamma^{2}}$ for two isolated atoms having the same energy level becomes exactly one, when electrodes electron kinetic energy  equate with energy level of the atom $\epsilon_{k}=\epsilon_{0}$.  See Fig. 3.
 Now in the presence of hybridization between energy orbital of these atoms, the normalized transmission probability $\dfrac{\Gamma^{2}}{(\epsilon_{k}-\epsilon_{a})^{2}+\Gamma^{2}}$ maxima shifts to molecular anti-bonding energy level $\epsilon_{k}=\epsilon_{a}=\epsilon_{0}+\tau$. In the absence of hybridization between energy orbital of the atoms, Green's function propagators interfere constructively and therefore the normalized transmission probability becomes exactly one, when electrodes electron kinetic energy equate with energy level of the atom $\epsilon_{k}=\epsilon_{0}$. The Green's function propagators for diatomic molecule interfere completely constructively for molecular anti-bonding state and completely destructively for molecular bonding state. This gives perfect transmission through molecular anti-bonding state and no transmission through molecular bonding state. Therefore, a single step in current and a single peak in differential conductance is shown up in Fig. 1.
 
 In Fig. 4 the normalized transmission probability for two atoms having different energy level is shown. As the kinetic energy of the electrodes electron equate with atom of the lower energy level half of the transmission $\dfrac{1}{2}$$\dfrac{\Gamma^{2}}{(\epsilon_{k}-\epsilon_{1})^{2}+\Gamma^{2}}$ through the device is achieved. While the next half of the transmission $\dfrac{1}{2}$$\dfrac{\Gamma^{2}}{(\epsilon_{k}-\epsilon_{2})^{2}+\Gamma^{2}}$ through the device is achieved where electrodes electron kinetic energy equate with atom of the higher energy level. In the presence of finite hybridization between energy orbitals of these atoms, molecular bonding and anti-bonding states are formed.
  Now the destructive interference of Green's function propagators for molecular bonding state suppresses the normalized transmission probability $\dfrac{1}{2}$$\dfrac{\Gamma^{2}}{(\epsilon_{k}-\epsilon_{a})^{2}+\Gamma^{2}}$$(1-\dfrac{2\tau}{\gamma})$, and the constructive interference of Green's function propagators for molecular anti-bonding state enhances the normalized transmission probability $\dfrac{1}{2}$$\dfrac{\Gamma^{2}}{(\epsilon_{k}-\epsilon_{a})^{2}+\Gamma^{2}}$$(1+\dfrac{2\tau}{\gamma})$.
   Therefore current step and differential conductance peak amplitude for molecular bonding state is smaller than for molecular anti-bonding state, as shown in Fig. 2. With increasing hybridization between energy orbital of these atoms, the asymmetry in normalized transmission probability, current, and differential conductance is increased.
\section{Conclusion}
In this study we have explicitly demonstrated  that current and differential conductance are suppressed for a diatomic molecular bonding state and consequently these are enhanced for anti-bonding state. The Green's function propagators entering into current and differential conductance calculations interfere destructively  for molecular bonding
state and constructively for anti-bonding state. Moreover for diatomic molecule comprising of two atoms of same energy level, no step in current and no peak in differential conductance has shown up for molecular bonding state. Therefore, for such a diatomic molecule current passes through molecular anti-bonding state and it does not pass through molecular bonding state. And for a diatomic molecule comprising of two atoms of different energy level, the current and differential conductance have shown to depend upon hybridization between energy orbital of two atoms. The more the orbital of the two atoms are hybridised the more current passes through molecular anti-bonding state and the less current passes through bonding state.

\end{document}